# An optical parsec-scale jet from a massive young star in the Large Magellanic Cloud

Anna F. McLeod[1*], Megan Reiter[2], Rolf Kuiper[3], Pamela D. Klaassen[4], Christopher J. Evans[4]

[1]School of Physical and Chemical Sciences, University of Canterbury, New Zealand
[2]Department of Astronomy, University of Michigan, Ann Arbor, USA
[3]Institute of Astronomy and Astrophysics, University of Tübingen, Germany
[4]UK Astronomy Technology Center, Royal Observatory Edinburgh, UK

**Highly collimated parsec-scale jets, generally linked to the presence of an accretion disk, are a commonly observed phenomenon from revealed low-mass young stellar objects[1,2]. In the past two decades, only a very few of these objects have been directly (or indirectly) observed towards high-mass (M > 8 M$_\odot$) young stellar objects[3,4,5,6,7], adding to the growing evidence that disk-mediated accretion is a phenomenon also occurring in high-mass stars[8,9,10,11], the formation mechanism of which is still poorly understood. Of the observed jets from massive young stars, none is in the optical regime (due to these being typically highly obscured by their native material), and none are found outside of the Milky Way. Here, we report the detection of HH 1177, the first extragalactic optical ionized jet originating from a massive young stellar object located in the Large Magellanic Cloud. The jet is highly collimated over the entire measured extent of at least 10 pc, and has a bipolar geometry. The presence of a jet indicates ongoing, disk-mediated accretion, and together with the high degree of collimation, this system is therefore likely to be an up-scaled version of low-mass star formation. We conclude that the physics governing jet launching and collimation is independent of stellar mass.**

At a distance of 50 kpc, the Large Magellanic Cloud (LMC) is an ideal search location for high-mass (M > 8 M$_\odot$) young stellar object (MYSO) jets: it is close enough to resolve such structures; it is nearly face-on, allowing a convenient viewing angle; it is undergoing active star formation; its low dust content makes it an environment in which the feedback effect of strong ionizing radiation is enhanced[12] and radiation pressure suppressed; it is a low-metallicity environment, allowing a resolved probe of star formation in conditions similar to those of earlier epochs of the Universe. In terms of MYSO jets, the LMC therefore allows the analysis of the effect of a different environment on the formation and propagation of jets in MYSOs, as current models are based on Milky Way observations. To date, only a handful of outflows from massive stars have been reported in the LMC[13,14], and all are detected in the radio regime with the Atacama Large Millimeter/submillimeter Array (ALMA). The driving sources of these outflows are still deeply embedded in their natal molecular cloud cores and the measured outflows are on sub-parsec scales with no optical jet counterparts.

The MYSO Herbig-Haro (HH) flow[15] reported here was detected in observations of the ionized star-forming region LMC N180 from the optical integral field spectrograph MUSE[16] at the Very Large Telescope (VLT). N180 consists of a classical, bubble-shaped HII region (see Figure 1) with a radius of ~2.5' (~36 pc). It is located on the border of a giant molecular cloud, it is currently undergoing star formation, and it hosts over 50 stars with masses M ≥ 10-15 M$_\odot$, as well as several early O-type stars[17].

The jet itself is externally ionized and detected as blue- and redshifted emission peaks of the Hα line, and spans a total of 10 pc. It is emerging from a pillar-like structure protruding from the southern rim of the surrounding ionized star-forming region. Its likely driving source is a MYSO of approximately 12 M$_\odot$ (corresponding to YSO N180-24 in [17]), which has formed at the tip of the pillar.

The jet-driving MYSO is optically visible at the origin of the blue and red lobes, which on its own corresponds to an unusual detection, as jet-driving MYSOs are usually not visible in the optical wavelength regime due to the high extinction caused by their embedding material. We suggest that the low dust content of the LMC has aided our investigation by enhancing the ionization feedback from the star on the surrounding matter and therefore allowing the situation in which the envelope has been dispersed before the disk, and the jet has escaped the pillar.

The medium spectral resolution and limited blueward coverage of the MUSE instrument does not allow the determination of a precise spectral type of the star, but together with the estimated mass of 12 $M_\odot$ and the absence of HeII absorption lines, we suggest that the driving source is an early-type B star (rather than a late O-type star). The jet is associated with HH emission traced by bow-shock structures where the jet lobes terminate (see Figure 1). The bow-shocks (which, together with the jet, make up the HH 1177 flow) are detected beyond the jet, and span 11 pc (0.76 arcminutes) from end to end, making this the second confirmed detection of an HH object beyond our own Galaxy[18], as well as one of the longest jets ever observed.

The detection of the jet not only as emission line peaks, but especially as a spatially resolved continuous structure traced by two distinct velocity components, indicates a coherent jet with a spatial orientation in which the red lobe is moving away and the blue lobe is moving towards the observer. The red lobe extends ~3.5 pc (0.24') to the Southeast of the pillar with a Position Angle of ~144 degrees. The first portion just below the source (at the position of aperture #4 in Figure 2) is partly obscured by the dominating pillar emission. It consists of a main segment and a bright terminal knot in close vicinity to the southern bow-shock. The blue lobe emerges from the top of the pillar with a Position Angle of ~-32 degrees, and spans ~6.5 pc (0.45'), although its terminal knot is confused by the projected vicinity of a star.

The jet is highly collimated along the entire detected parsec-scale extent. With an apparent half-width of approximately $R = 0.1$ pc (0.4") throughout, the lower limits for length-to-half-width ratios are 35 and 65 for the red and blue lobes, respectively. We note that the jet width is an upper limit, as the true width is unresolved in our observations, and the length-to-half-width is therefore a strict lower limit (see Methods section). The measured radial velocities corresponding to the offsets in wavelength of the red and blue peaks are of the order of 300-400 km/s, yielding a (non de-projected) dynamic timescale of the jet in the range of $2.8 \cdot 10^4 - 3.7 \cdot 10^4$ yr, indicating that despite the short lifetime of massive stars, the accretion and jet phase is at least of the order of some 10 kyr. Figure 2 shows the continuum-subtracted spectra of the red and blue lobes: the red and blue peaks are clearly identified right and left of the central H$\alpha$ line. What is seemingly a red peak in the blue spectrum corresponds to an OH sky line doublet[19] at 6577.183 and 6577.3863 Å.

From the intensity of the red and blue emission line peaks and the estimated width of the jet body, we compute the mass-loss rate of the jet to be $\sim 2.9 \cdot 10^{-6}$ $M_\odot$ yr$^{-1}$, and a resulting mass accretion rate of $\sim 9.5 \cdot 10^{-6}$ $M_\odot$ yr$^{-1}$. The mass-loss and accretion rates obtained for the jet are comparable to literature values for embedded MYSOs[20], confirming the high-mass nature of the driving source. Furthermore, these values also confirm that the source is young (given that the mass-loss and accretion rates are expected to decrease with the age of the star[20]), strengthening our result of an optical jet from an unobscured MYSO.

In terms of the total (projected) length of ~11 pc of HH 1177, the jet, together with the detected bow shocks, is comparable to some of the largest HH flows observed in the Milky Way[3,21,22]. However, traditional HH objects are emission-line nebulosities corresponding to shock-excited gas along a protostellar outflow; direct detection of the jet body is possible with external ionization[23,24] in massive star-forming regions. In the case of HH 1177, the jet itself is detected, allowing the direct measurement of the length-to-(half)width ratio, extent and velocity, as well as a description of its morphology and spatial orientation. The high degree of collimation and the measured radial velocities found for HH 1177 are similar to those found in lower-mass stars[25,26], therefore supporting our theoretical knowledge about the connection of accretion disks and magneto-centrifugal ejection as well as the launching and collimation physics of jets: the underlying processes of jet launching and collimation are independent of stellar mass[4,27,28]. In terms of the accretion-ejection link, the detection of this large-scale jet implies active disk accretion epochs on a regular basis, which denote a necessity in the formation of massive stars to circumvent the radiation pressure problem[10].

**Acknowledgments**
R. K. acknowledges financial support within the Emmy Noether Research Program funded by the German Research Foundation (DFG) under grant no. KU 2849/3-1.

**Author contributions**
A. F. M is the principal investigator of the MUSE observations used for this work. A. F. M. reduced and analyzed the data, and wrote the initial manuscript. R. K. provided the input concerning the theoretical aspect of the interpretation; C. J. E. analyzed the stellar spectrum to determine a first spectral classification. M. R. and P. D. K. provided essential input concerning YSO jet and pillar observations and jet mass-loss rates. All coauthors commented on the manuscript.

**Author Information**
Correspondence and requests for materials should be addressed to A. F. M. (anna.mcleod@canterbury.ac.nz).

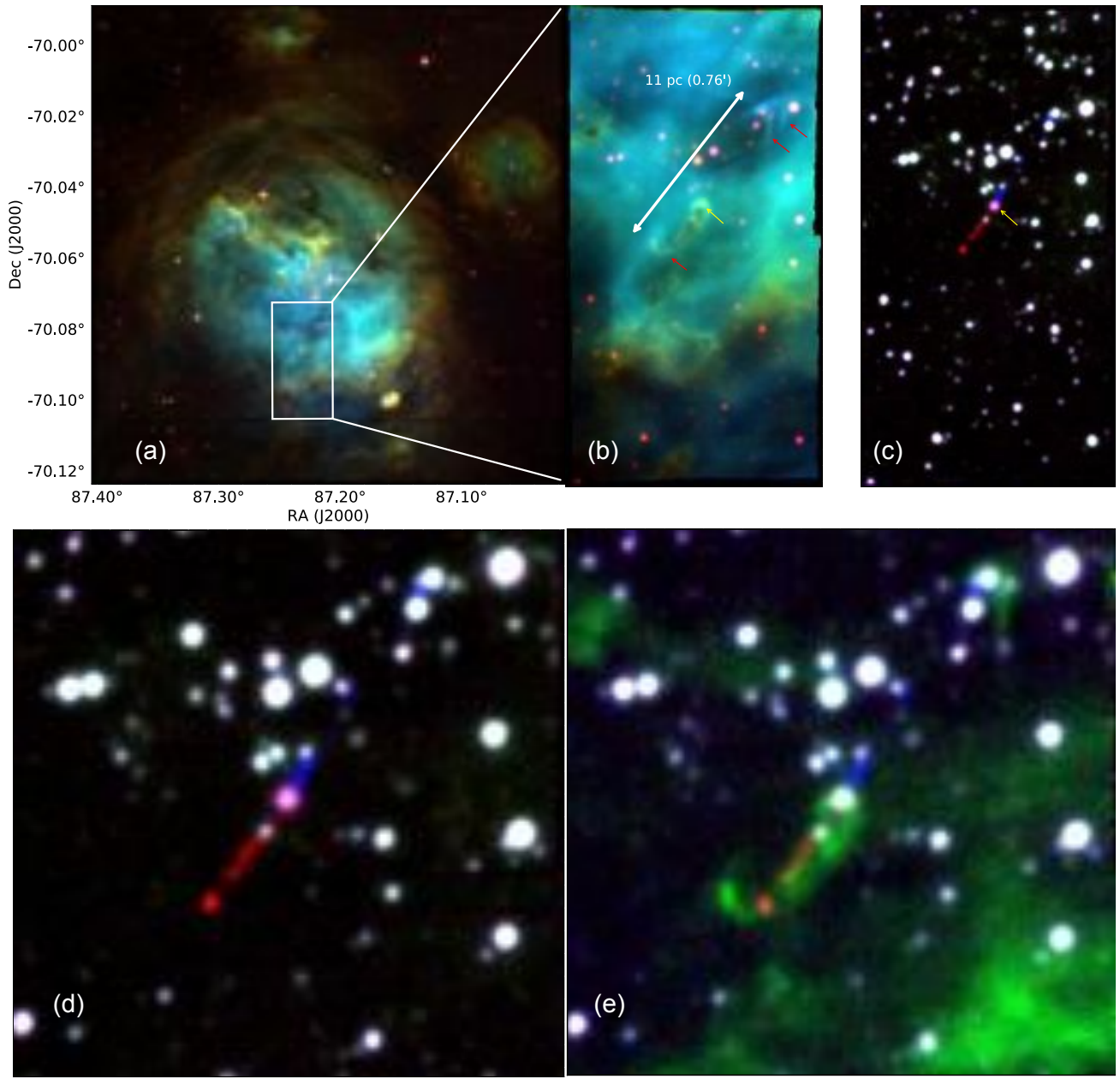

***Figure 1. RGB composites of the star-forming region N180 and the jet.*** *Panels (a) and (b): three-color composites of LMC N180 (red = [SII]6731, green = Hα, blue = [OIII]5007). Panel (b): the red arrows point at the bow-shocks, the yellow arrow indicates the jet source. Panel (c): three-color composite of the same region as (b), where the red and the blue correspond to the red and blue Hα emission line peaks, and the green corresponds to the image of the collapsed MUSE data cube. The RGB in (d) is the same as in (a), in (e) green corresponds to [SII]6731. In all figures: North is up and East is left.*

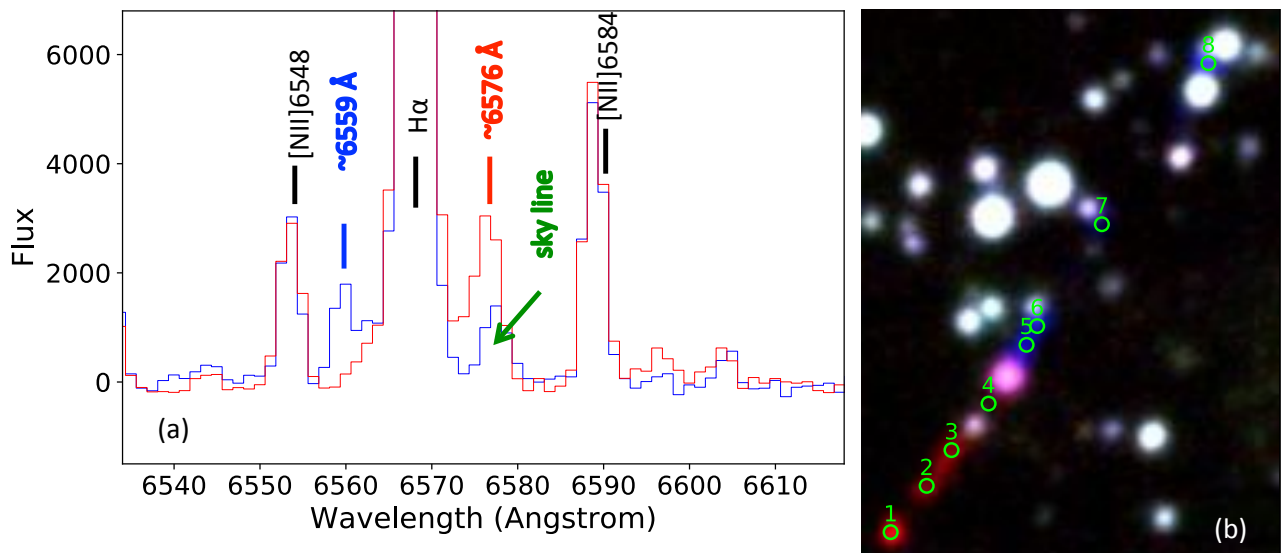

*Figure 2. **Spectrum of the red and blue jet lobes.*** *Panel (a): co-added spectra of the red and blue jet lobes (accordingly color-coded), extracted from circular apertures centered on the green circles (#1 for the red, #5 for the blue) in (b). The flux is measured in 10-20 erg/Å/s/cm2. Estimated levels of noise in the spectra ~92 and 129 for the red and blue, respectively. The strong, central Hα emission is coming from the HII region, as is evident from the diffuse emission in Figure 1b, where green is the central Hα wavelength: it does not include the red and blue emission line peaks, and the jet is not visible.*

## METHODS

### Optical integral field spectroscopy

The data used for this analysis is optical integral field data from the MUSE instrument[16], mounted on the Very Large Telescope in Chile, as part of the observing program 096.C-0137(A) (PI McLeod). MUSE has a field of view of about 1x1 arcminutes and a pixel scale of 0.2 arcseconds. The data was observed between November and December 2015, and reduced with the MUSE data reduction pipeline[29] in the *Esorex* environment, using the available standard calibrations for each night. The entire mosaic of the region N180 consists of 64 single pointings, each observed twice with a 90-degree rotation dither pattern and an exposure time of 90 seconds. The pointings were observed in the MUSE nominal wavelength range (4650-9300 Å, R = 2000-4000) and the Wide Field Mode field-of-view. From the calibrated MUSE data cube, we estimate the noise of the spectra from the continuum flux closest to the analyzed lines (in the range from 6610 Å to 6660 Å).

### Integrated line maps and spectral fitting

The Hα, [SII]6731 and [OIII]5007 integrated line maps used to produce the three-color image of Figure 1 were obtained with the Python package Spectral Cube[30] as for MUSE data of other pillar-like objects[31,32]. The jet is not visible in the central Hα image, and it is only by navigating through the individual slices in the data cube in the region around the Hα line, that the jet becomes evident. We therefore produced integrated line maps as for the main emission lines, centered on the red and blue emission respectively.

To determine the amplitude and central wavelength of the blue- and redshifted peaks of the Hα line, which trace the two lobes of the bipolar jet, we used the Python package Pyspeckit[33] to fit multiple Gaussian components to the emission lines in the region covering the [NII] and the Hα lines. Since the MUSE spectra span approximately 4600 Å in wavelength, they cover almost all of the nebular emission lines typically found in HII regions as well as a multitude of sky lines, the latter being particularly dense in the red part of the spectra. Given the multitude of lines, identifying and fitting the continuum across the entire covered MUSE wavelength range can be quite challenging and may result in an overestimation of the continuum, resulting in a negative baseline after subtraction. We therefore crop the spectra to a smaller portion in the range from 6000 to 6680 Å as to allow a more precise estimate of the local continuum. The fitted spectra are shown in Extended Data Figures 1 and 2. The fit to the blue lobe in apertures #5 and #7 does not appear optimal: adding a fifth Gaussian component centered between the blue and the Hα yields what seems to be better fit by eye for the blue peak, but the resulting values are within the errors of the 4-component fit, and we therefore do not include an additional component. The best fit parameters are summarized in Extended Data Tables 1 and 2. Error estimates are taken from the Pyspeckit fitting routine, and correspond to the RMS of the residuals. For each spectrum shown, we also state the level of noise estimated from the continuum closest to the analyzed lines, as described above.

### Jet widths and collimation

We measure the (projected) diameter of the two lobes by extracting slices across the jet structure perpendicular to the jet direction at positions #1, #2, #3, #5, #6 and #8 in Figure 2b (position #4 corresponds to the part of the red lobe covered by the emission from the pillar, and position #7 is contaminated by the emission from a nearby star). We estimate the diameter $L$ of the jet body (and the half-width $R = L/2$) at each position by again fitting Gaussians to the obtained profiles (as shown in Extended Data Figure 3). These widths were then compared to the jet length shown in Figure 1 (see main text). However, due to a seeing-limited spatial resolution of 0.6" (0.15 pc) at the distance of the LMC, the true width of the jet is unresolved and the values obtained from the described fit are strictly upper limits. As a consequence, the resulting length-to-half-width ratios are a strict lower limit.

### Mass-loss and accretion rate

We compute the mass-loss rate by first computing the electron density $n_e$ from the measured intensity $I_{H\alpha,red}$ and $I_{H\alpha,blue}$ of the emission line peaks (in units of $10^{-15}$ erg/cm$^2$/s/arcsec$^2$), and then combining this with the measured radial velocities $V$ and width $L_{pc}$ of the jet body in pc[34]:

$$n_e = 15.0 \, (I_{H\alpha}/L_{pc})^{1/2} \, \text{cm}^{-3} \quad (1)$$

$$\dot{M}_{jet} = \mu m_H n_e V \pi (L/2)^2 f \quad (2)$$

where $\mu$ is the mean molecular weight for which we assume a value of 1.35, $m_H$ is the proton mass, $V$ the jet velocity and $f$ a geometric filling factor inferred from the morphology of the images. We measure $I_{H\alpha,red} \cong 0.224 \pm 0.001$ (in units of $10^{-15}$ erg/cm$^2$/s/arcsec$^2$) from a circular aperture on the terminal red knot (centered on aperture #1) in the integrated intensity map, and use $L \cong (0.08 \pm 0.04)$ pc to obtain $n_e \cong (25.13 \pm 0.01)$ cm$^{-3}$. Since the width of the jet is not resolved in the MUSE data, we use 0.08 pc as the jet width, which is comparable to HH80-81 (from Figure 1 in [21]). For this assumed jet width we adopt a large uncertainty of ±50%. Together with a radial velocity of $V \cong (363.79 \pm 0.46)$ km/s and $f = 1$, we obtain a mass-loss rate of $(1.57 \pm 0.99) \cdot 10^{-6}$ M$_\odot$ yr$^{-1}$. Similarly, for the blue lobe, we obtain $I_{H\alpha,blue} \cong 0.139 \pm 0.001$ from aperture #5, and a mass-loss rate of $(1.28 \pm 0.81) \cdot 10^{-6}$ M$_\odot$ yr$^{-1}$, with $V \cong (376.56 \pm 0.46)$ km/s. Finally, we compute the combined mass-loss rate to be ~ $2.85 \cdot 10^{-6}$ M$_\odot$ yr$^{-1}$. Assuming a ratio between the jet mass-loss rate and the accretion rate of ~0.3[35,36], we obtain a mass accretion rate of ~$9.5 \cdot 10^{-6}$ M$_\odot$ yr$^{-1}$.

**Data availability statement**

The dataset that supports the findings of this study is available in the ESO archive (http://archive.eso.org/eso/eso_archive_main.html) under the observing program 096.C-0137(A) executed at the VLT (ESO Paranal, Chile). Additional requests can be directed to the corresponding author.

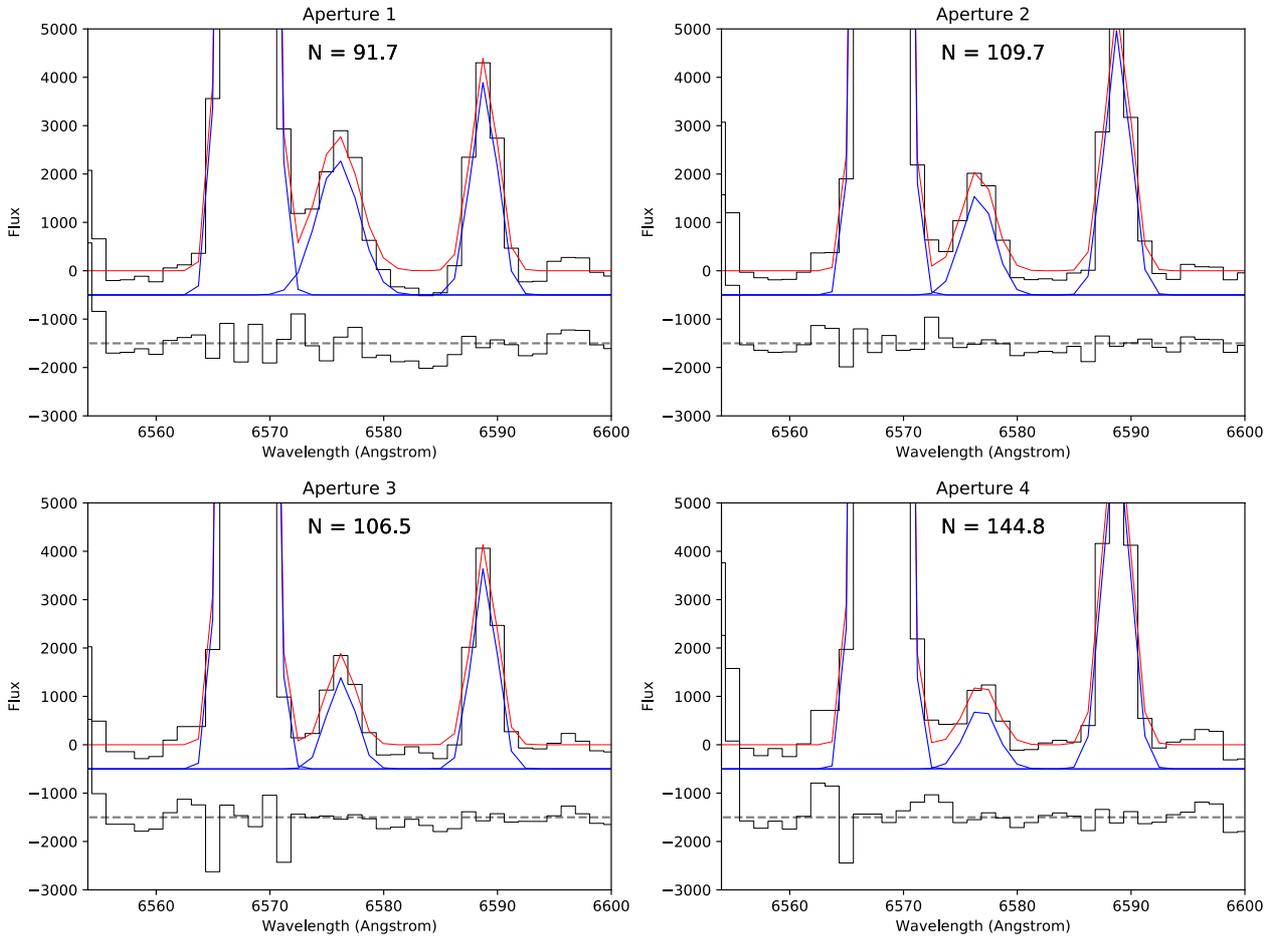

***Extended Data Figure 1***. ***Fitted spectra of the red jet lobe.*** *Co-added spectra of the red lobe of the jet (extracted from a 2 pixel radius circular aperture centered on the green circles in Figure 2b, main text), continuum-subtracted and fitted with a 3-component Gaussian. The spectrum is shown in black, the fit in red, the single components are plotted in blue and the residuals of the fit are shown below each spectrum (the latter two are shown with an offset on the y-axis for better display). The estimated noise is stated for each spectrum. Best-fit parameters are summarized in Extended Data Table 1. The flux is in units of $10^{-20}$ erg/s/Å/cm$^2$.*

| | Hα | | | Red peak | | | [NII] | | |
|---|---|---|---|---|---|---|---|---|---|
| Ap. # | Peak Flux | Centroid | Width | Peak Flux | Centroid | Width | Peak Flux | Centroid | Width |
| 1 | 82399.0 ± 416.8 | 6568.0 ± 7.5×10$^{-3}$ | 1.2 ± 7.7×10$^{-3}$ | 3057.0 ± 331.0 | 6575.9 ± 0.3 | 2.1 ± 0.3 | 4649.5 ± 430.6 | 6588.8 ± 0.1 | 1.1 ± 0.1 |
| 2 | 99262.3 ± 570.9 | 6568.1 ± 7.5×10$^{-3}$ | 1.15 ± 7.7×10$^{-3}$ | 2184.6 ± 466.8 | 6576.5 ± 0.4 | 1.7 ± 0.4 | 5604.3 ± 554.9 | 6588.8 ± 0.1 | 1.2 ± 0.1 |
| 3 | 80343.2 ± 429.0 | 6568.0 ± 7.2×10$^{-3}$ | 1.2 ± 7.4×10$^{-3}$ | 2027.6 ± 390.9 | 6576.3 ± 0.3 | 1.4 ± 0.3 | 4281.4 ± 431.3 | 6588.8 ± 0.1 | 1.1 ± 0.1 |
| 4 | 140227.0 ± 735.8 | 6568.0 ± 6.5×10$^{-3}$ | 1.1 ± 6.8×10$^{-3}$ | 1132.7 ± 485.0 | 6576.2 ± 1.2 | 2.4 ± 1.2 | 7338.1 ± 691.4 | 6588.7 ± 0.1 | 1.2 ± 0.1 |

***Extended Data Table 1***. ***Best fit parameters of the Gaussian fitting to the red lobe***. *The spectra and fits are shown in Extended Data Figure 1. Fluxes are in units of $10^{-20}$ erg/s/Å/cm$^2$, centroids and widths are in Å.*

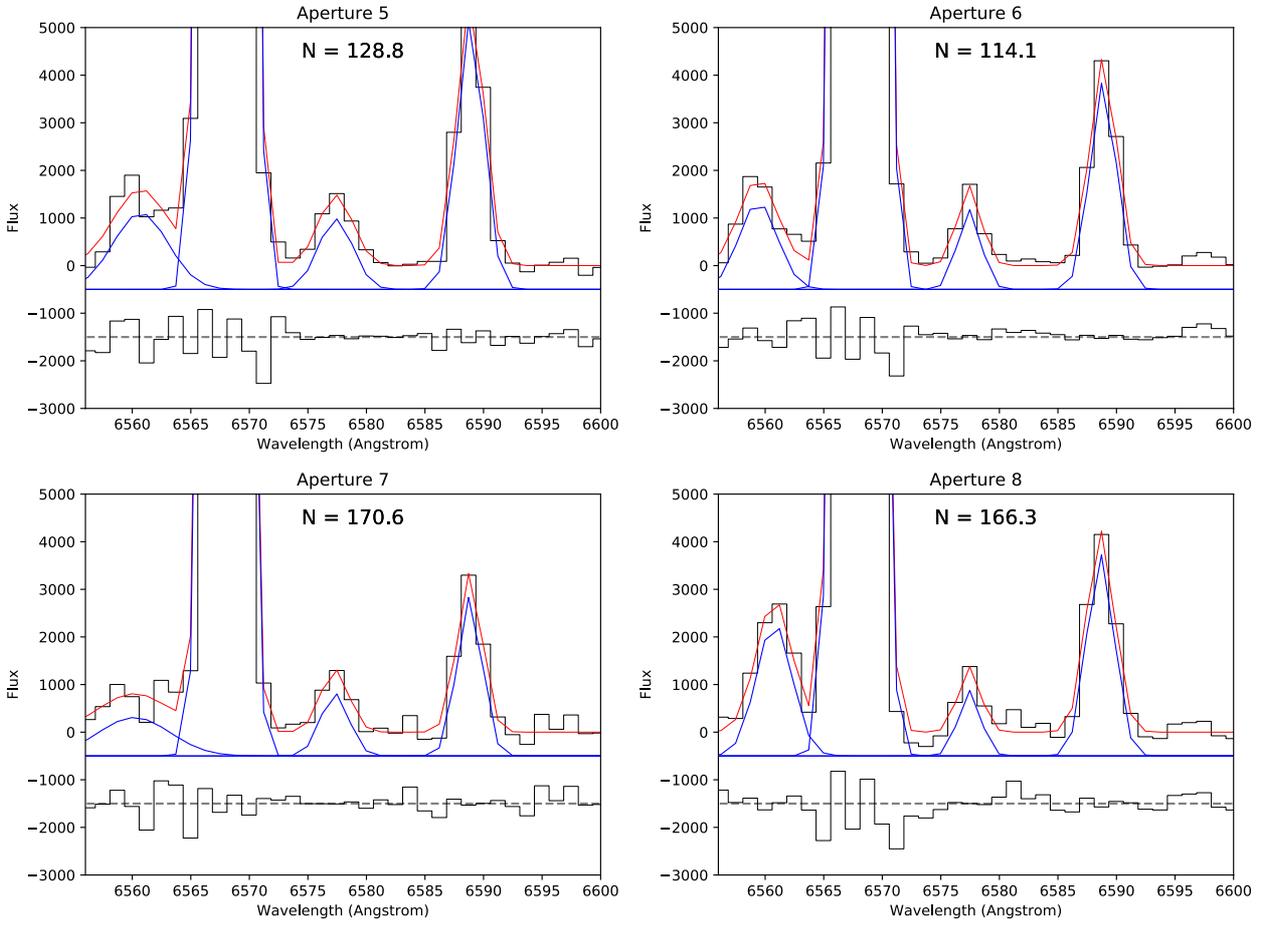

***Extended Data Figure 2***. ***Fitted spectra of the blue jet lobe.*** *Co-added spectra of the blue lobe of the jet (extracted from a 2 pixel radius circular aperture centered on the green circles in Figure 2b, main text), continuum-subtracted and fitted with a 4-component Gaussian. The spectrum is shown in black, the fit in red, the single components are plotted in blue and the residuals of the fit are shown below each spectrum (the latter two are shown with an offset on the y-axis for better display). The estimated noise is stated for each spectrum. Best fit parameters are summarized in Extended Data Table 2. The flux is in units of $10^{-20}$ erg/s/Å/cm$^2$.*

| | **Blue** | | | **Hα** | | | **Sky** | | |
|---|---|---|---|---|---|---|---|---|---|
| Ap. # | Peak Flux | Centroid | Width | Peak Flux | Centroid | Width | Peak Flux | Centroid | Width |
| 5 | 1606.5 ± 203.2 | 6560.7 ± 0.3 | 2.3 ± 0.4 | 159788.0 ± 295.7 | 6568.1 ± 2.4×10$^{-3}$ | 1.1 ± 2.5×10$^{-3}$ | 1484.4 ± 251.8 | 6577.4 ± 0.3 | 1.5 ± 0.3 |
| 6 | 1834.4 ± 210.6 | 6559.4 ± 0.2 | 1.6 ± 0.2 | 126890.0 ± 258.1 | 6568.1 ± 2.6×10$^{-3}$ | 1.1 ± 2.6×10$^{-3}$ | 1677.1± 257.4 | 6577.4 ± 0.2 | 1.0 ± 0.2 |
| 7 | 808.1 ± 153.9 | 6560.2 ± 0.7 | 3.1 ± 0.9 | 85319.3 ± 249.3 | 6568.0 ± 3.7×10$^{-3}$ | 1.1 ± 4.0×10$^{-3}$ | 1317.6 ± 229.2 | 6577.3 ± 0.2 | 1.2 ± 0.2 |
| 8 | 2792.7 ± 261.5 | 6560.8 ± 0.2 | 1.5 ± 0.2 | 79684.5 ± 300.5 | 6567.9 ± 5.0×10$^{-3}$ | 1.2 ± 5.1×10$^{-3}$ | 1378.0 ± 310.8 | 6577.5 ± 0.3 | 0.9 ± 0.2 |

***Extended Data Table 2***. ***Best fit parameters of the Gaussian fitting to the blue lobe.*** *The spectra and fits are shown in Extended Data Figure. 2. Fluxes are in units of $10^{-20}$ erg/s/Å/cm$^2$, centroids and widths are in Å (the parameters of the [NII] line are not reported in this table, as not directly relevant).*

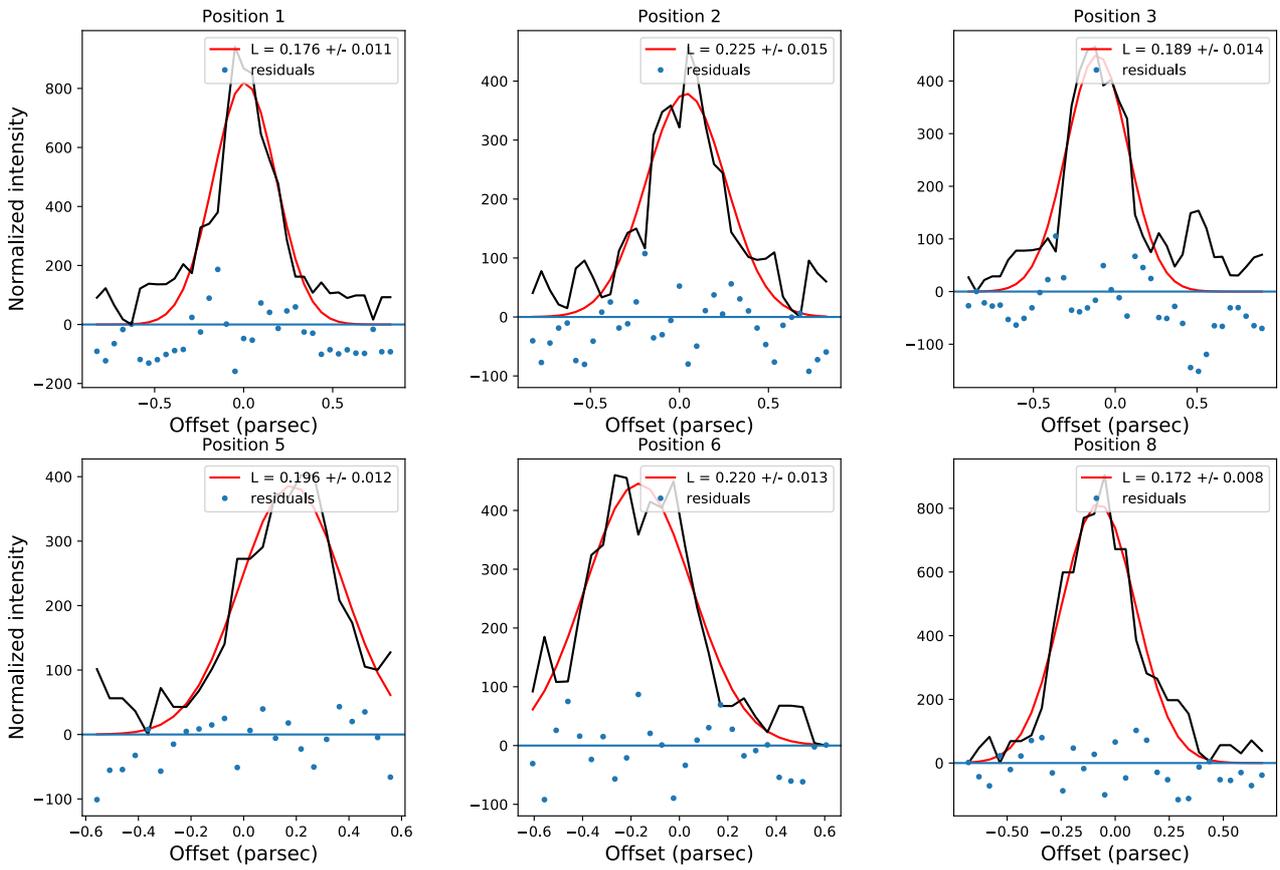

***Extended Data Figure 3. Diameter of the jet body.*** *Black curves: integrated line map intensity profiles along virtual slits positioned perpendicular to the jet axis on the positions marked in Figure 2b of the red (positions #1, 2 and 3) and on the blue lobes (positions #5, 6 and 8). Red curves: best fit Gaussian profiles; blue diamonds: residuals.*